\documentclass{WileyMSP-template}

\usepackage{amsmath}
\usepackage{amssymb}

\usepackage{cite}

\usepackage[labelfont=bf]{caption}

\newcommand{\beq}{\begin{equation}}
\newcommand{\enq}{\end{equation}}

\newcommand{\omr}{\ensuremath{\omega_\mathrm{r}}}

\newcommand{\gbar}{\ensuremath{\bar{\gamma}_\mathrm{T}}}
\newcommand{\gint}{\ensuremath{\gamma_\mathrm{x}}}
\newcommand{\Nint}{\ensuremath{N_\mathrm{x}}}
\newcommand{\gtr}{\ensuremath{\gamma_\mathrm{tr}}}
\newcommand{\Ntr}{\ensuremath{N_\mathrm{tr}}}

\begin{document}

\pagestyle{fancy}
\rhead{\includegraphics[width=2.5cm]{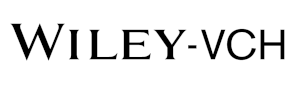}}

\title{Recent Developments in Quantum-Circuit Refrigeration}

\maketitle


\author{Timm Fabian Mörstedt*},
\author{Arto Viitanen*},
\author{Vasilii Vadimov},
\author{Vasilii Sevriuk},
\author{Matti Partanen},
\author{Eric Hyyppä},
\author{Gianluigi Catelani},
\author{Matti Silveri},
\author{Kuan Yen Tan},
\author{Mikko Möttönen}



\begin{affiliations}
Timm Fabian Mörstedt, Arto Viitanen, Dr. Vasilii Vadimov, Prof. Mikko Möttönen \\
QCD Labs, QTF Centre of Excellence, Department of Applied Physics \\
Aalto University\\
00076 AALTO, Finland\\
E-mail:
timm.morstedt@aalto.fi, arto.viitanen@aalto.fi\\
Dr. Vasilii Sevriuk, Dr. Matti Partanen, Eric Hyyppä, Dr. Kuan Yen Tan \\
IQM \\
02150 Espoo, Finland \\
Dr. Gianluigi Catelani \\
JARA Institute for Quantum Information (PGI-11) \\
Forschungszentrum J{\"u}lich \\
52425 J{\"u}lich, Germany \\
Quantum Research Centre \\
Technology Innovation Institute \\
Abu Dhabi, UAE \\
Dr. Matti Silveri \\
Nano and Molecular Systems Research Unit \\
University of Oulu \\ 
90014 Oulu, Finland 

\end{affiliations}



\begin{abstract}
We review the recent progress in direct active cooling of the quantum-electric degrees freedom in engineered circuits, or quantum-circuit refrigeration. In 2017, the invention of a quantum-circuit refrigerator (QCR) based on photon-assisted tunneling of quasiparticles through a normal-metal--insulator--superconductor junction inspired  a series of experimental studies demonstrating the following main properties: (i) the direct-current (dc) bias voltage of the junction can change the QCR-induced damping rate of a superconducting microwave resonator by orders of magnitude and give rise to non-trivial Lamb shifts, (ii) the damping rate can be controlled in nanosecond time scales, and (iii) the dc bias can be replaced by a microwave excitation, the amplitude of which controls the induced damping rate. Theoretically, it is predicted that state-of-the-art superconducting resonators and qubits can be reset with an infidelity lower than $10^{-4}$ in tens of nanoseconds using experimentally feasible parameters. A QCR-equipped resonator has also been demonstrated as an incoherent photon source with an output temperature above one kelvin yet operating at millikelvin. This source has been used to calibrate cryogenic amplification chains. In the future, the QCR may be experimentally used to quickly reset superconducting qubits, and hence assist in the great challenge of building a practical quantum computer.

\end{abstract}


\section{Introduction}
Superconducting quantum circuits~\cite{Nakamura1999,Wallraff2004, Krantz2019, Wallraff2021} currently constitute one of the most promising platforms for quantum information processing and quantum simulations. A superconducting quantum computer has been shown to achieve quantum supremacy~\cite{Arute2019}, an important step toward large-scale fault-tolerant quantum computation~\cite{Preskill2018}. However, fast and precise initialization of qubits, which is required for example for efficient quantum error correction, still remains a challenge. This is particularly important for multi-qubit processors, in which the fidelity of the initialization of the whole system decays exponentially with the number of qubits, provided the fidelity of a single qubit initialization is fixed. A naive method of qubit initialization would be cooling down the system to millikelvin temperatures and waiting until it decays to its ground state. However, the characteristic lifetimes of the modern superconducting qubits reach values of tens to hundreds of microseconds~\cite{Rigetti2012,riste_millisecond_2013,yan_flux_2016, wang_transmon2021}, which makes such an approach impractical in terms of performance. Consequently, there is need for active reset protocols which are inherently free of this drawback. 

\begin{figure}[h]
\centering
  \includegraphics[width=0.9\linewidth]{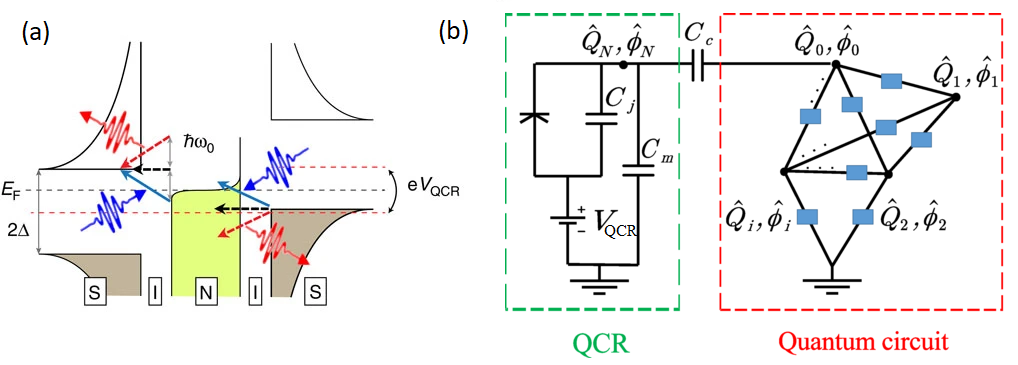}
  \caption{a) Energy diagram of photon-assisted tunneling in a QCR. The blue arrows show the tunneling processes accompanied by single-photon absorption from the refrigerated circuit, while the red arrows correspond to tunneling with single-photon emission to the circuit. Here, $\Delta$ is the superconductor gap parameter and $V_{\mathrm{QCR}}$ is the bias voltage across the SINIS junction. At the bias voltage illustrated in the figure, the photon emission events as well as elastic tunneling (black arrows) are energetically forbidden (dashed arrows), owing to the gap in the density of states of the superconductor and the lack of high-energy electron or hole excitations in the normal metal. Reproduced with permission \cite{Tan2017}. Copyright 2017, Springer Nature. b) Lumped-element diagram for the QCR capacitively coupled to a generic quantum circuit, where $\hat{Q}_k$ denotes the $k$:th node charge opertor and $\hat{\phi}_k$ its conjugate phase. The blue boxes may represent capacitances, inductances, Josephson junctions, or parallel combinations of them. The green-dashed side of the diagram shows the voltage biased NIS junction which represents the QCR. Reproduced with permission \cite{Hsu2020}. Copyright 2020, American Physical Society.}
  \label{fig:QCR_circuit}
\end{figure}

Many of the existing active initialization protocols are quite system-specific~\cite{Valenzuela2006, Reed2010, Bultink2016, CampagneIbarcq2013, Mariantoni2011, Geerlings2013, Zhou2021} and operate only on the lowest or a couple of lowest excited levels of the system, which potentially renders them vulnerable to leakage errors out of the computational subspace. The relatively recent concept of a quantum circuit refrigerator (QCR)~\cite{Tan2017,Silveri2017}, which constitutes a voltage-controlled broadband environment for quantum systems, allows to overcome the above-mentioned issues of speed, accuracy, and high-lying excitations. 

The QCR consists of two normal-metal--insulator--superconductor (NIS) junctions, biased by an external voltage, where the normal-metal island is capacitively coupled to the system to be refrigerated. In the course of electron tunneling between the superconducting leads and the normal-metal island, the QCR can absorb and emit photons from or to the rest of the circuit and, thus, effectively change its temperature, see \textbf{Figure~\ref{fig:QCR_circuit}}. At low bias voltages $|e V_\mathrm{QCR}| < 2 \Delta$, where $\Delta$ is the superconductor-specific gap parameter, photon absorption is dominant over emission due to the Pauli principle and the energy gap in the density of states of the superconductor. Hence, the QCR is prone to act as a low-temperature environment for the circuit, down to environmental temperature half of the electron temperature in the normal metal. The energy gap in the superconducting density of states and the low electron temperature render the characteristic decay rate induced by the QCR on the coupled quantum circuit tunable by orders of magnitude by the voltage between the superconducting leads of the QCR. Thus although at vanishing bias the QCR is essentially decoupled from the system, at bias voltages close to $2 \Delta/e$, the quantum-circuit refrigeration can be carried out within a very short time, down to single nanoseconds~\cite{Sevriuk2019,Hsu2020,Hsu2021}.

Another advantage of the QCR is its versatility. It can be used for refrigeration of various systems, including qubits~\cite{Hsu2020,Yoshioka2021}, resonators~\cite{Silveri2017, Tan2017, Sevriuk2019}, arbitrary linear and non-linear many-element systems as in Figure~\ref{fig:QCR_circuit}, and even distributed systems.

In this paper, we review the latest experimental and theoretical results on the QCR. In Sec.~\ref{sec2}, we feature experiments on quantum-limited heat transfer by microwave photons over macroscopic distances. These experiments led to the invention of the QCR although photon-assisted tunneling was not significant. The following sections describe a QCR coupled to a resonator, including resonator reset (Sec.~\ref{sec3}), Lamb shift  (Sec.~\ref{sec4}), quickly tunable dissipation (Sec.~\ref{sec5}), the related exceptional points (Sec.~\ref{sec6}), and QCR operation in the radio-frequency regime (Sec.~\ref{sec7}). Furthermore, we study the high-voltage regime of a QCR in Sec.~\ref{sec8} and~\ref{sec9} for photon generation and amplification. In Sec.~\ref{sec10}, we present the theoretical background of the QCR, whereas Sec.~\ref{sec11} provides a summary and outlook of the research.

\section{Quantum-limited Heat Conduction}\label{sec2}
\begin{figure}[h]
\centering
  \includegraphics[width=0.5\linewidth]{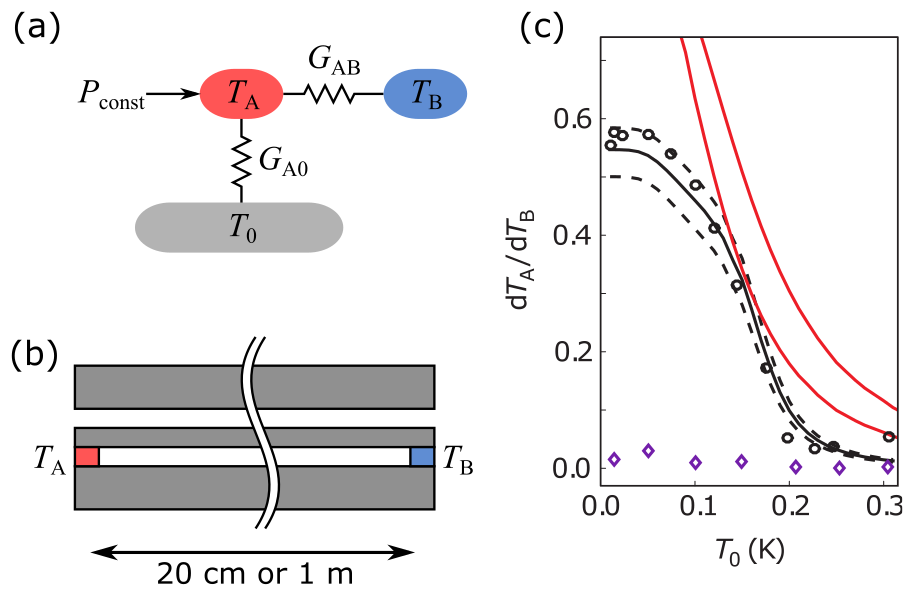}
  \caption{ Investigation of quantum-limited thermal conductance. 
  a) Simplified thermal model. The thermal conductance $G_\textrm{AB}$ between two normal-metal islands at electron temperatures $T_\textrm{A}$ and $T_\textrm{B}$ can ideally reach the quantum of thermal conductance. 
  In the experiment, $T_\textrm{B}$ is swept, which affects $T_\textrm{A}$ and enables extracting $G_\textrm{AB}$.
  In addition, the electrons in island~A are coupled to the phonon bath at temperature $T_0$ via conductance $G_\textrm{A0}$. 
  Constant heat flow from higher temperatures is denoted by $P_\textrm{const}$.
  b) Schematic presentation of the coplanar waveguide transmission line that is terminated by the resistive normal-metal islands A and B. 
  c) Differential temperature response $\textrm{d}T_\textrm{A}/\textrm{d}T_\textrm{B}$ as a function of the phonon bath temperature $T_0$. The experimental data from the main device is denoted by black circles. 
  The red curves represent the ideal case  $G_\textrm{AB}=G_\textrm{Q}$, with two different literature values for the electron--phonon coupling. 
  Black solid and dashed lines show the result of a more detailed theoretical model detailed in Ref.~\cite{partanen_quantum-limited_2016}. 
  The purple diamonds denote a control device, in which the photonic channel through the waveguide is suppressed by shorting the center conductor to ground with superconducting leads. Adapted with permission \cite{partanen_quantum-limited_2016}. Copyright 2016, Springer nature.}
  \label{fig:Q_limited}
\end{figure}

In this section, we discuss the utilization of elastic electron tunneling through NIS junctions in studying quantum-limited heat conduction in a superconducting microwave transmission line. These results have been reported in Ref.~\cite{partanen_quantum-limited_2016}.
Importantly,  elastic tunneling through NIS junctions can be used as a refrigerator and heater for the electron excitations in the normal-metal electrode~\cite{giazotto_opportunities_2006}. 
Furthermore, NIS junctions can also serve as an \emph{in situ} thermometer probing the electron temperature in the normal metal~\cite{giazotto_opportunities_2006}.
The experimental device consists of a superconducting coplanar waveguide transmission line that is connected to ground potential at both ends via resistors, or normal-metal islands, see schematic in \textbf{Figure~\ref{fig:Q_limited}}b.
The length of the transmission line is 20~cm or 1~m in the experiment and it has a spiral structure.

In contrast to the experiments discussed in the sections below, here photon-assisted, also referred to as inelastic tunneling, is of vanishing importance owing to the good impedance matching of the terminating resistors. Thus instead of a QCR environment, the normal-metal islands appear simply as Ohmic resistors to the microwaves in the transmission line. In fact, impedance-matched resistors at finite electron temperature act as one-dimensional black-body radiators, ideally absorbing all incoming photons and emitting more photons to the transmission line the higher their temperature. These mechanisms give rise to heat exchange between the resistors via the transmission line. 

The simplified thermal model of the system is presented in Figure~\ref{fig:Q_limited}a. 
Notably, island~A at temperature $T_\textrm{A}$ is in thermal contact with a phonon bath at temperature $T_0$ due to electron--phonon coupling ($G_\textrm{A0}$).
Moreover, island~A can exchange heat with island~B at temperature $T_\textrm{B}$ through the photonic channel over the transmission line ($G_\textrm{AB}$).
In the ideal case, the photonic heat conduction reaches the theoretical upper limit for a single-channel heat conduction, the quantum of thermal conductance [$G_\textrm{AB} \approx G_\textrm{Q} = \pi k_\textrm{B} T_0/(6\hbar)$] for small temperature differences $T_0 \approx T_\textrm{A} \approx T_\textrm{B}$.
In addition, a constant small heat flow with a power $P_\textrm{const}$ reaches island~A.
In the steady state, the total heat flow into and out of island~A must be equal. 
Hence, one obtains an equation for the differential temperature dependence $dT_\textrm{A}/dT_\textrm{B} = 1/(1+aT_0^3)$, where $a$ is a predetermined constant that only depends on the material and geometry of the normal metal.
This differential temperature dependence increases with decreasing bath temperature, which is a clear indication of heat flow through the photonic channel since it has a weaker temperature dependence than electron--phonon coupling.
The experimental results confirm this increase implying a very efficient photonic heat transfer, see  Figure~\ref{fig:Q_limited}c.

The quantum of thermal conductance is a statistics-independent universal upper bound for single-channel heat conduction~\cite{pendry_quantum_1983,rego1999,schmidt_photon-mediated_2004}, and has been observed in various systems~\cite{Schwab2000,meschke_single-mode_2006,timofeev_electronic_2009,Jezouin2013,Cui2017,Banerjee2017}.
The experiment discussed here \cite{partanen_quantum-limited_2016} was the first one over a macroscopic distance.

\section{Invention of the Quantum-Circuit Refrigerator}
\label{sec3}

In a similar fashion to coupling resistors to a microwave transmission line as in Sec.~\ref{sec2}, they can be utilized to cool coplanar waveguide (CPW) resonators. Namely, a low-temperature resistor dissipates photons from a coupled resonator. However, such a bare resistive environment is not tunable. Fortunately, in an effort to realize such a resistive environment in 2017, Tan et al.~\cite{Tan2017} observed an unexpectedly high cooling power arising, not from Ohmic losses, but from photon-assisted tunneling illustrated in Figure~\ref{fig:QCR_circuit}a. This observation led to the invention of the QCR. Let us describe the first experiments in more detail below.

\begin{figure}[h]
  \includegraphics[width=\linewidth]{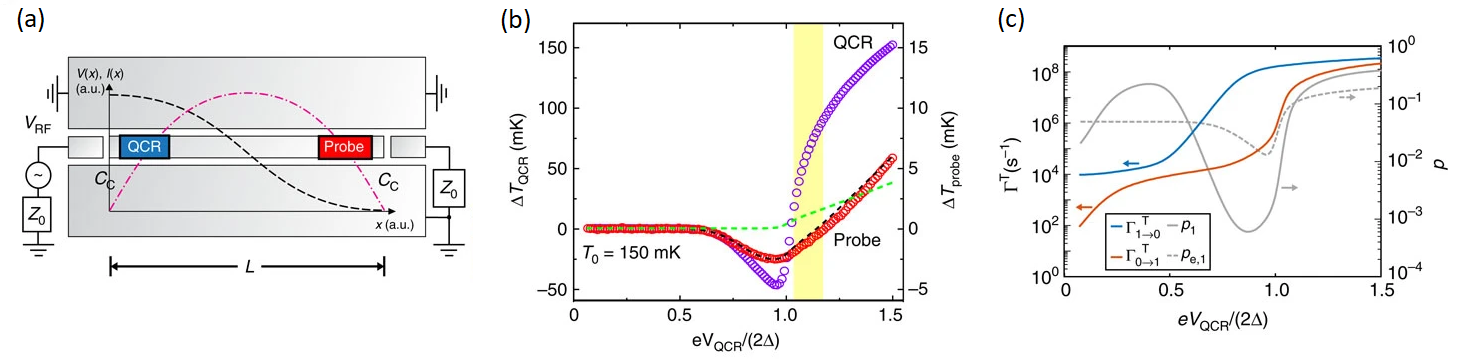}
  \caption{First observation of quantum-circuit refrigeration. a) Sample design with a QCR and a probe resistor embedded at opposite ends of a co-planar waveguide resonator. The dotted lines indicate current (red) and voltage (black) of the fundamental mode. The experimental setup also allows for an external microwave drive $V_{\mathrm{RF}}$. b) Electron temperature change $\Delta T$ as a function of the QCR bias voltage, measured at the QCR (purple circles) and at the probe resistor (red circles). Theoretical curves are shown  including (black) and excluding (green) photon-assisted tunneling. c) Transition rates $\Gamma_{0\rightarrow 1}$ (excitation) and $\Gamma_{1\rightarrow 0}$ (relaxation) obtained from $P(E)$ theory using parameters corresponding to the measured sample. The steady-state photon occupation follows $p_{1}=\Gamma_{0\rightarrow 1}/(\Gamma_{0\rightarrow 1}-\Gamma_{1\rightarrow 0})$ for the full model (grey line) and for the approximation of equal resonator and QCR normal-metal temperature (grey dashed line). Reproduced with permission\cite{Tan2017}. Copyright 2017, 
  Springer Nature.}
  \label{fig:Tan}
\end{figure}

In this first QCR experiment, both the QCR and a probe resistor are embedded into the resonator, see \textbf{Figure~\ref{fig:Tan}}a. The resistors at close but not at the ends of the resonator to allow for significant Ohmic dissipation of resonator photons at the resistors. Instead of only two NIS junctions at the QCR, there is an additional pair of NIS junctions which are used to monitor the electron temperature of the island at different QCR bias voltages. Identical NIS thermometry is utilized at the probe resistor to find out its electron temperature. 

Figure~\ref{fig:Tan}b shows the main result of the QCR experiment, i.e., QCR and probe electron temperatures at functions of the QCR bias voltage. Both the QCR and probe reach their minimum electron temperatures just below $eV_{\mathrm{QCR}}=2\Delta$. At slightly higher bias voltages, the QCR electron temperature rises quickly above its zero-bias value owing to heating of its electron cloud by elastic tunneling. The Ohmic coupling of the QCR to the resonator tends thus to heat up the resonator mode, but since the photon-assisted tunneling still cools the resonator more strongly than the Ohmic coupling heats it, the net effect is cooling the resonator. This is observed at the probe resistor side as electron temperature lower than that at zero bias. The fact that the probe electrons are cooled down while the QCR electrons are heated up lead to the need to find an additional heat conduction mechanism to those considered in Sec.~\ref{sec2}, i.e., photon-assisted tunneling which is also referred to as quantum-circuit refrigeration in this context.

Figure~\ref{fig:Tan}b also shows the result of a thermal model where the QCR electron temperature is given as an input. The electron temperature of the probe resistor is in good agreement with theory, provided that the contributions from photon-assisted tunneling is included. Without photon-assisted tunneling, there is no significant cooling at the probe side. To further improve cooling power, several QCRs can potentially be stacked as a cascade cooler~\cite{Camarasa2014} or connected to low-temperature Ohmic baths~\cite{partanen_quantum-limited_2016}.\\

When the QCR bias voltage exceeds $2\Delta$ as shown in the yellow region in Figure~\ref{fig:Tan}b, the electron temperature at the QCR increases rapidly. However, the probe temperature at the resonator is not immediately affected. Strategic QCR placement close to the endpoint of the resonator with minimum mode current leads to negligible Ohmic coupling to the resonator, minimizing Ohmic heating from the QCR electrons. Therefore, such a design enables optimal cooling close to the $2\Delta$ threshold without excessive dissipation and sensitivity to rapid heating even when slightly exceeding this gap voltage~\cite{Tan2017}.\\

Although we discuss below the contemporary way to model the QCD based on a systematic first-principles perturbation theory~\cite{Silveri2017}, the first theoretical model for the quantum-circuit refrigerator was based on a so-called $P(E)$ theory~\cite{Ingold1992}, the predictions of which for the QCR-induced resonator relaxation and excitation rates for the measured device are shown in Figure~\ref{fig:Tan}c. We observe that the bias voltage applied to the QCR can change the transition rates $\Gamma_{0\rightarrow 1}$ (excitation) and $\Gamma_{1\rightarrow 0}$ (relaxation) of the fundamental resonator mode by orders of magnitude. As a result, the steady-state excitation probability $p_{1}=\Gamma_{0\rightarrow 1}/(\Gamma_{0\rightarrow 1}-\Gamma_{1\rightarrow 0})$ can be tuned by up to three orders of magnitude within the range $|eV_{\mathrm{QCR}}|<2\Delta$. In particular, the minimum value $p_{1}<10^{-3}$ is reached just below $eV_{\mathrm{QCR}}=2\Delta$, in a similar range that also shows most efficient cooling. Since these properties lead to a fast decay of the system to its quantum ground state, the QCR is a promising candidate for controlled reset of quantum circuits, most notably superconducting qubits. The initialization rate is competitive with those of other current qubit initialization protocols~\cite{Reed2010, Bultink2016, CampagneIbarcq2013, Mariantoni2011, Geerlings2013, Zhou2021}.

\section{Broadband Lamb Shift}
\label{sec4}

Physical quantum systems are inevitably open to their environment, which among other effects, shifts the energy levels of the system. Even the case of coupling to an environment in the vacuum state may significantly perturb the system. In this case, the shift caused by the vacuum, modelled as broadband electromagnetic vacuum fluctuations, is known as the Lamb shift, first observed in an atomic system by Lamb et al. in 1947~\cite{Lamb47}. The once-considered-negligible shift has become important since the emerging engineered quantum systems call for extreme precision of energy levels, such as resonator and qubit frequencies, of the order of the Lamb shift. An incomplete understanding of the frequencies may lead to unintended dissipation or cross-coupling in a quantum circuit, disturbing the operation of the quantum device in question.

\begin{figure}[h]
  \centering
  \includegraphics{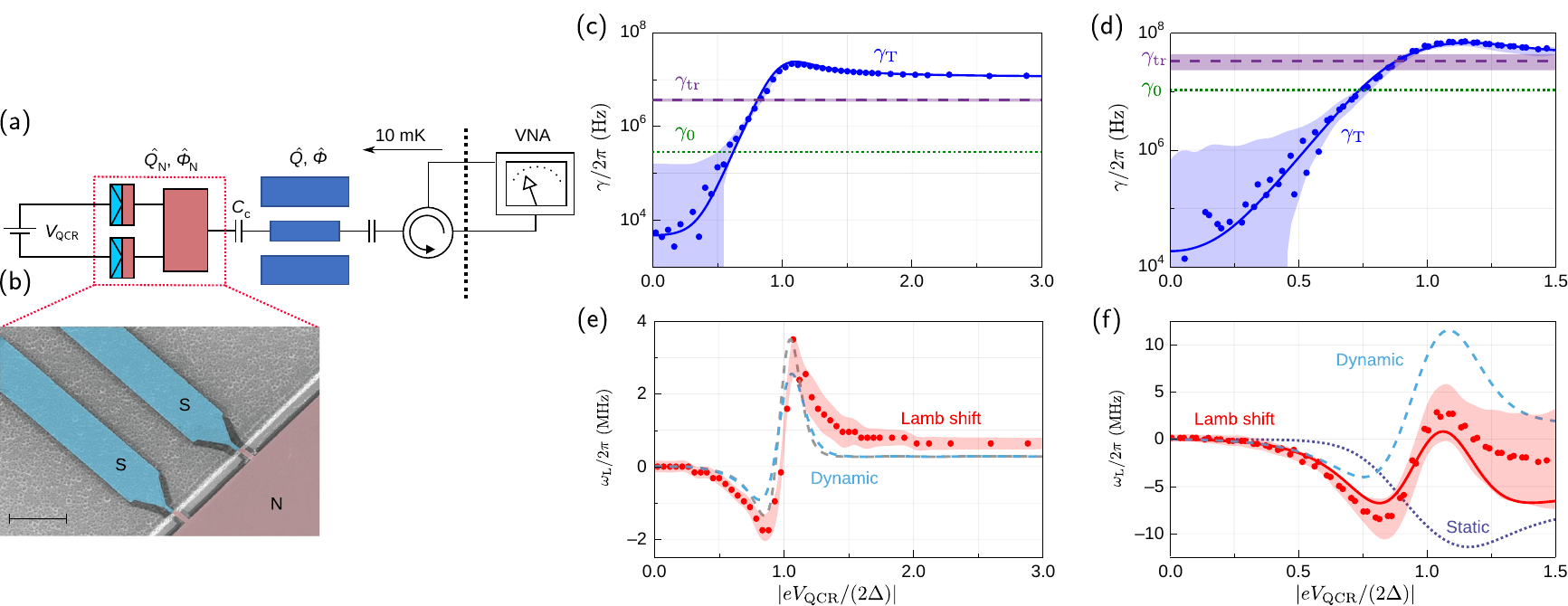}
  \caption{Lamb shift. a) Schematic illustration of the device and measurement setup. A CPW resonator is capacitively coupled to the normal-metal lead of a voltage biased QCR and to a transmission line used for readout. b) False-colour scanning electron micrograph of the NIS junctions. The scale bar is $5~\mu$m wide. c,d) Resonator coupling strength $\gamma_\mathrm{T}$ to the essentially vacuum environment induced by the QCR and e,f) the resulting Lamb shift $\omega_\mathrm{L}$ of the resonator frequency as functions of the bias voltage applied to the QCR. Dots denote experimental data and solid lines model prediction. The horizontal lines indicate coupling strengths to the transmission line $\gamma_\mathrm{tr}$ and to excess sources $\gamma_\mathrm{0}$. The shaded regions denote $1\sigma$ confidence intervals. We show data for two different samples, for one with a fundamental resonance frequency $4.67$~GHz in panels c) and e) and for the other with $8.54$~GHz in panels d) and f). See Ref.~\cite{Silveri19} for details. Adapted with permission~\cite{Silveri19}. Copyright 2019, Springer Nature.}
  \label{fig:lambshift}
\end{figure}

The Lamb shift can be measured in an engineered quantum system by coupling a QCR to a superconducting CPW resonator, as shown in \textbf{Figure~\ref{fig:lambshift}}a,b. In this scheme where a single QCR is coupled at one end of a half-wave-length resonator, the QCR effectively provides a bath of harmonic oscillators essentially in vacuum as well as the possibility to control the coupling strength $\gamma_\mathrm{T}$ between this environment and the resonator by approximately three orders of magnitude. Thus, the Lamb shift $\omega_\mathrm{L}$ of the resonator mode with bare angular frequency $\omega_\mathrm{r}$ can be controlled with the bias voltage of the QCR and measured in a typical microwave reflection experiment of the resonator mode as discovered by Silveri et al. in 2019~\cite{Silveri19}. The key results of these experiments, i.e., the coupling strength and the Lamb shift as functions of the QCR bias voltage are shown in Figure~\ref{fig:lambshift}c--f.

Note that a quantum circuit may also exhibit other energy shifts, most notably an ac Stark shift, which shifts the energy levels of anharmonic systems due to the excitation of its environment. Although critical for the operation of qubits, a harmonic oscillator does not experience it, thus providing an ideal platform for measuring the Lamb shift in an engineered quantum circuit.

The Lamb shift of the resonator mode caused by the broadband vacuum environment is characterized by the coupling strength $\gamma_\mathrm{T}$ and obtained with second-order perturbation theory as~\cite{Silveri19}
\begin{equation} \label{eq:lambshift}
    \omega_{\mathrm{L}}=-\mathrm{PV} \int_{0}^{\infty} \frac{\mathrm{d} \omega}{2 \pi}\left[\frac{\gamma_{\mathrm{T}}(\omega)}{\omega-\omega_{\mathrm{r}}}+\frac{\gamma_{\mathrm{T}}(\omega)}{\omega+\omega_{\mathrm{r}}}-2 \frac{\gamma_{\mathrm{T}}(\omega)}{\omega}\right],
\end{equation}
where PV denotes principal value integration. The integration takes the environment into account as a continuum of modes with angular frequency $\omega$, each with a corresponding coupling strength to the harmonic oscillator $\gamma_\mathrm{T}(\omega)$. The coupling strength is obtained from the photon-assisted electron-tunneling theory of a QCR developed in Ref.~\cite{Silveri2017}. The first two parts in the integral are induced by the photon-assisted tunnelling processes, while the resonator-frequency-independent term arises from elastic tunneling. Notably, the elastic processes do not affect the coupling strength nor the effective temperature, yet they promote a Lamb shift. At high biases, $eV_\mathrm{QCR}/(2\Delta)\gtrsim2$, the electromagnetic environment becomes Ohmic and the Lamb shift vanishes. Note however that in the experiment above, the frequency shift is measured with respect to zero bias, which artificially yields a non-vanishing shift at high bias.

\section{Quickly Tunable Dissipative Environment for Resonators}
\label{sec5}
The ability to tune the decay rate of the CPW resonator mode described above~\cite{Tan2017, Silveri19} suggests that the QCR can be utilized to initialize superconducting quantum-information circuits and quantum-information systems coupled to such circuits~\cite{kurizki_quantum_2015}. It has also been theoretically predicted~\cite{Hsu2020,Hsu2021} that a QCR coupled to the quantum-information system can have a negligible effect in the QCR off-state, and can convert it to a highly damped system in the QCR on-state. However, it is also important to demonstrate precise and rapid control over the QCR in order to use it for fast qubit reset, as well as to show high reset fidelity in such a regime. The first experimental work in this direction is described in Ref.~\cite{Sevriuk2019} which shows rapid photon evacuation from the CPW resonator by a short voltage pulse applied to the QCR junctions. 

The system used in Ref.~\cite{Sevriuk2019} is a similar design as above in Figure~\ref{fig:lambshift}a,b and contains the QCR which is capacitively coupled to one end of a half-wave-length CPW resonator. At the other end, the resonator is capacitively coupled to an rf transmission line. Both superconducting electrodes of the QCR are connected to transmission lines. 

\begin{figure}[h]
  \includegraphics[width=0.8\linewidth]{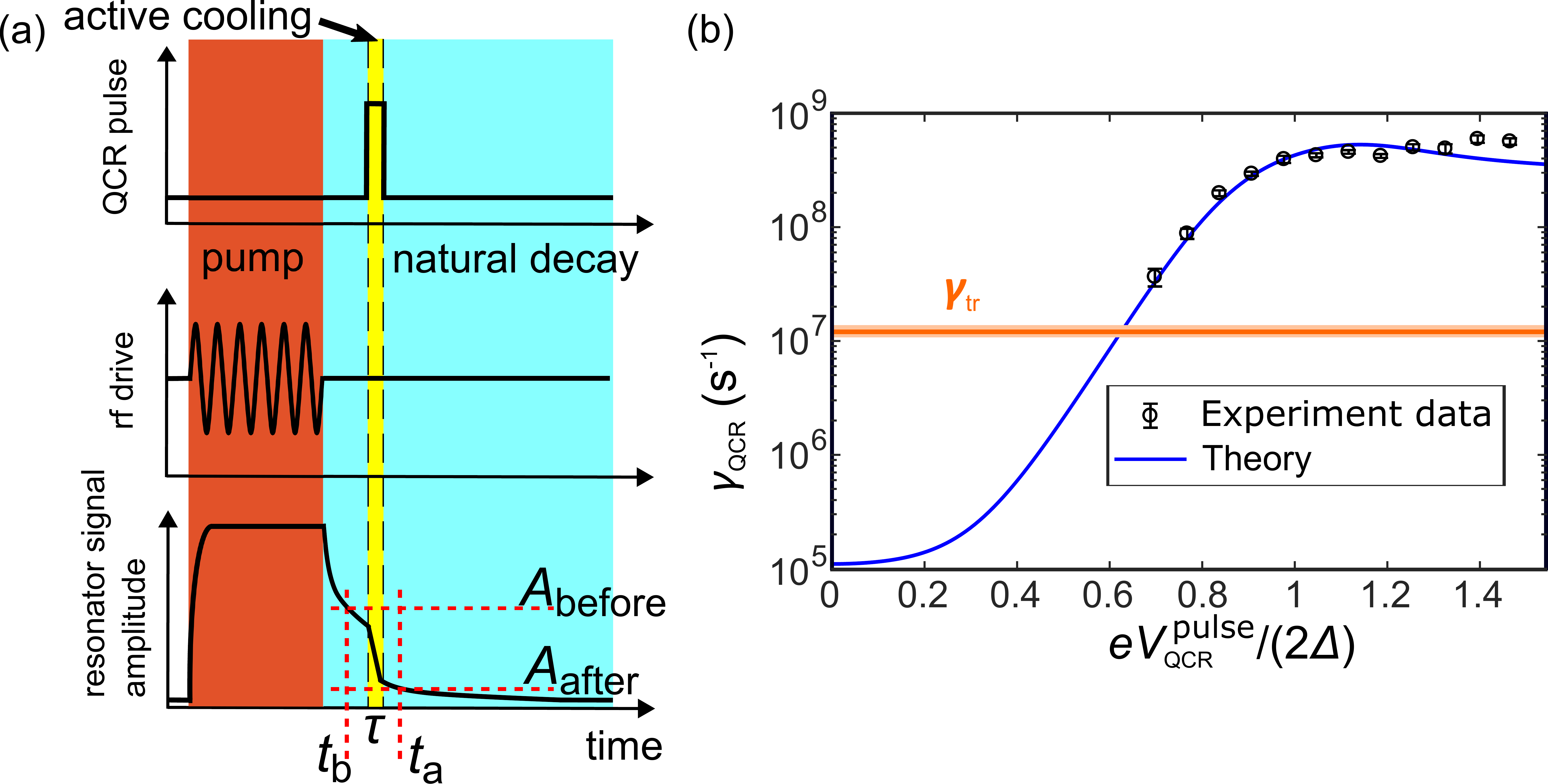}
  \caption{
  \label{fig:pulse_qcr_result} Fast control of dissipation using a QCR. a) Pulsing scheme of the experiment. During the time interval indicated by red color, the resonator is driven with a resonant rf tone. Then the drive is switched off and the decay of the resonator signal is observed (light blue region). Yellow color indicates the time period when a voltage pulse with width $\tau$ is sent to the QCR to induce active cooling. To extract the damping rate of the QCR, we vary the pulse length between two chosen points in time, before ($t_\mathrm{b}$) and after ($t_\mathrm{a}$) the QCR pulse and detect the corresponding drop in the resonator signal. b) Resonator damping rate by the QCR $\gamma_{\mathrm{QCR}}$ as a function of the amplitude of the QCR voltage pulse. We normalize the amplitude by the energy gap of the superconducting leads, which is $2\Delta=2\times215$~$\mathrm{\mu}$eV. The orange line with the shaded region denotes the damping rate of the transmission line $\gamma_{\mathrm{tr}}$ together with its 1$\sigma$ uncertainty. Adapted with permission\cite{Sevriuk2019}. Copyright 2019, AIP Publishing.}
\end{figure}

The basic experimental protocol is described in \textbf{Figure~\ref{fig:pulse_qcr_result}}a. First, we send a microwave pump pulse of $8.863\,\mathrm{GHz}$ through the rf transmission line to populate the fundamental mode of the resonator. Subsequently, we observe the natural decay of the signal using the same transmission line. At a chosen instant of time, a short voltage pulse is applied to the QCR bias. During this pulse, the QCR induces a change in the decay rate of the signal, the magnitude of which depends on the pulse amplitude. Due to the fact that the QCR pulse lengths are in the range of 5--20~ns it is not possible to accurately fit an exponentially decaying function to the measured voltage signal during the pulse. Consequently, the decay rate during the pulse is extracted by varying the pulse length and fitting the observed signal drop after the pulse as a function of the pulse length. This method also helps to remove the effect of the pulse rise and fall edges. An example of the experimental result is shown in Figure~\ref{fig:pulse_qcr_result}b. A flat top pulse with a Gaussian rise and fall edges with 170~ps length was used in this particular example.

As shown in Figure~\ref{fig:pulse_qcr_result}, the data agrees well with the photon-assisted-tunneling theory, which is based on the parameters of the sample defined by fabrication and obtained by dc QCR measurements. In the figure, we also observe the resonator decay rate due to its coupling to the transmission line, which dominates the decay in the QCR off-state. It also prevents us to directly measure the QCR-induced decay rate below the $10^7$ s$^{-1}$ range. The theoretical results indicate that the QCR can change the decay rate of the coupled circuit by roughly four orders of magnitude. In this experiment, it has been demonstrated that the resonator decay rate can be rapidly changed by a factor of 55 during the voltage pulse applied to the QCR. Based on this result, we conclude that with the optimal pulse amplitude, the photon number in the resonator can be reduced to less than 1\% of the original number in a few tens of nanoseconds, which seems advantageous in comparison to other reset schemes~\cite{Reed2010, Bultink2016, CampagneIbarcq2013, Mariantoni2011, Geerlings2013, Zhou2021}.

Further work is in progress to demonstrate also high-fidelity reset of highly coherent single qubits. This effort is supported by the theory recently developed in Refs.~\cite{Hsu2020,Hsu2021}.

\section{Realization of Exceptional Points in Superconducting Circuits}
\label{sec6}
\begin{figure}[h]
  \includegraphics[width=0.8\linewidth]{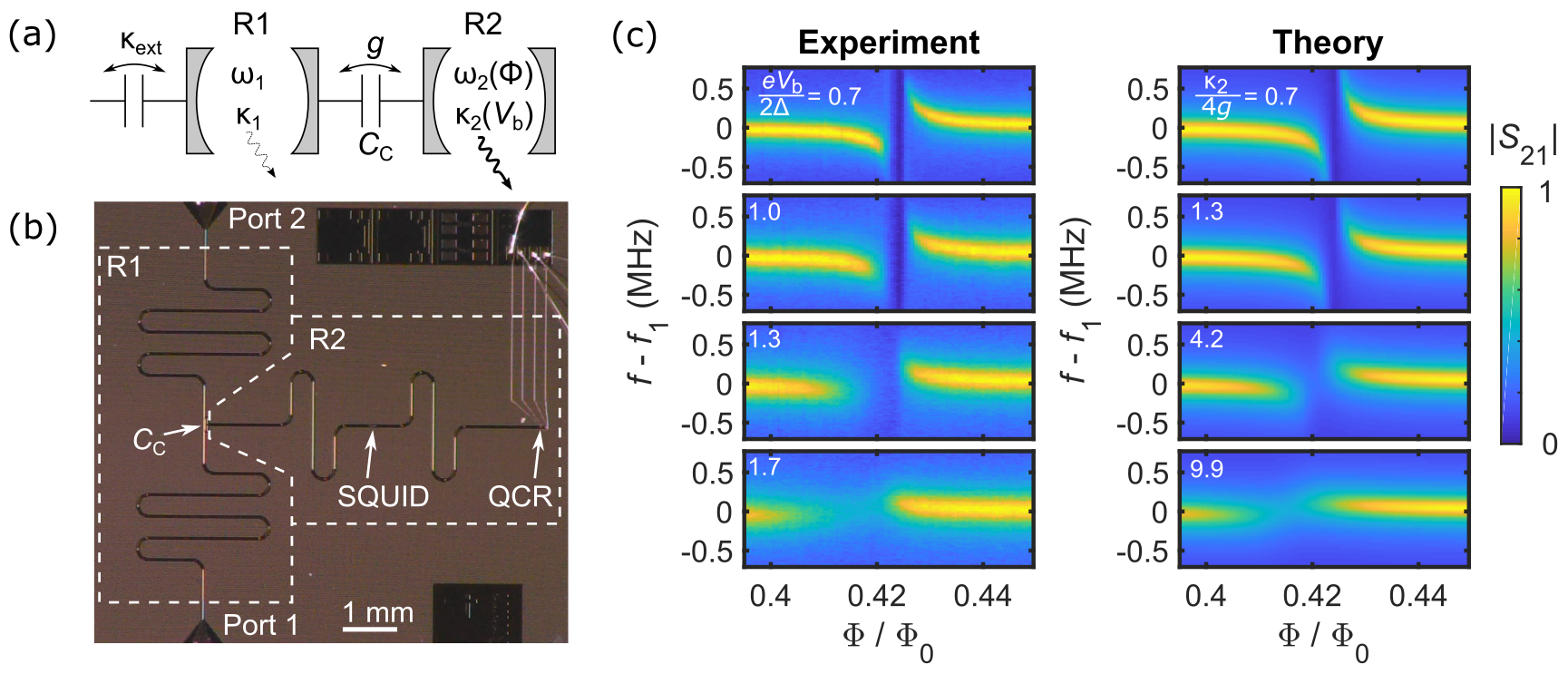}
  \caption{Realization of an exceptional point in a superconducting microwave circuit~\cite{partanen_exceptional_2019}.
  a) Schematic representation  of the device consisting of two resonators R1 and R2 with angular frequencies $\omega_i$ and dissipation rates $\kappa_i$, $i=1,2$.
  b) Optical micrograph of the device. 
  The resonator R2 contains a SQUID and a QCR for frequency and dissipation tunability, respectively.
  c) Experimental and theoretical transmission coefficient $|S_{21}|$ as a function of the magnetic flux $\Phi$ through the SQUID and the probe frequency $f$. 
  Here, $\Phi_0$ denotes the flux quantum, and $f_1 = 5.223$ GHz is the resonance frequency of R1, when R2 is far-detuned.
  We show experimental data at different bias voltages $V_\textrm{QCR}$ accross the QCR as indicated, and match those with theoretical results obtained for different dissipation rates $\kappa_2$.
  Adapted with permission\cite{partanen_exceptional_2019}. Copyright 2019, American Physical Society.}
  \label{fig:EP}
\end{figure}

In recent years, the study of non-Hermitian systems has gained substantial attention owing to their rich physical phenomena. 
In particular, the focus is on square-root singularities, or so-called exceptional points, that are characterized by coalescing eigenvalues~\cite{kato1966perturbation, Heiss2012, Xu2016,Doppler2016,Ding2016,Ding2018,El-Ganainy2018} and eigenvectors of an effective system Hamiltonian.
One can realize such a system in superconducting circuits using a QCR as a voltage-tunable source of dissipation and a SQUID for frequency tuning, as reported in Ref.~\cite{partanen_exceptional_2019}.
The studied system is shown in \textbf{Figure~\ref{fig:EP}}a,b.
The circuit consists of two capacitively coupled (capacitance $C_\mathrm{C}$, coupling strength $g$) resonators: resonator~R1 has a low dissipation rate $\kappa_1$ and a fixed angular frequency $\omega_1$, and resonator~R2 has  a tunable dissipation $\kappa_2$ and a tunable angular frequency $\omega_2$. 

The system can be described by an effective Hamiltonian 
\begin{equation}
    H =  \begin{pmatrix}
  -\textrm{i} \kappa_1 /2 & g  \\
  g & \delta - \textrm{i}\kappa_2/2
 \end{pmatrix},
\end{equation}
where the detuning is $\delta = \omega_2 - \omega_1 $.
By tuning the resonators into resonance, $\delta = 0$, and adjusting the loss rate $\kappa_2 = 4 g$, the system reaches the exceptional point.
Here, it is assumed that $\kappa_2 \gg \kappa_1$.
Experimentally, the exceptional point is the transition point between avoided crossing of resonance frequencies and a single resonance frequency with flux modulation. 
This transition is clearly demonstrated in Figure~\ref{fig:EP}c. 
Notably, the exceptional point enables optimally efficient heat flow between the resonators, i.e., there is no back and forth oscillation of energy but ideal exponential decay instead.

\section{Radio-Frequency Quantum-Circuit Refrigerator}
\label{sec7}
\begin{figure}[h]
  \centering
  \includegraphics[width=\linewidth]{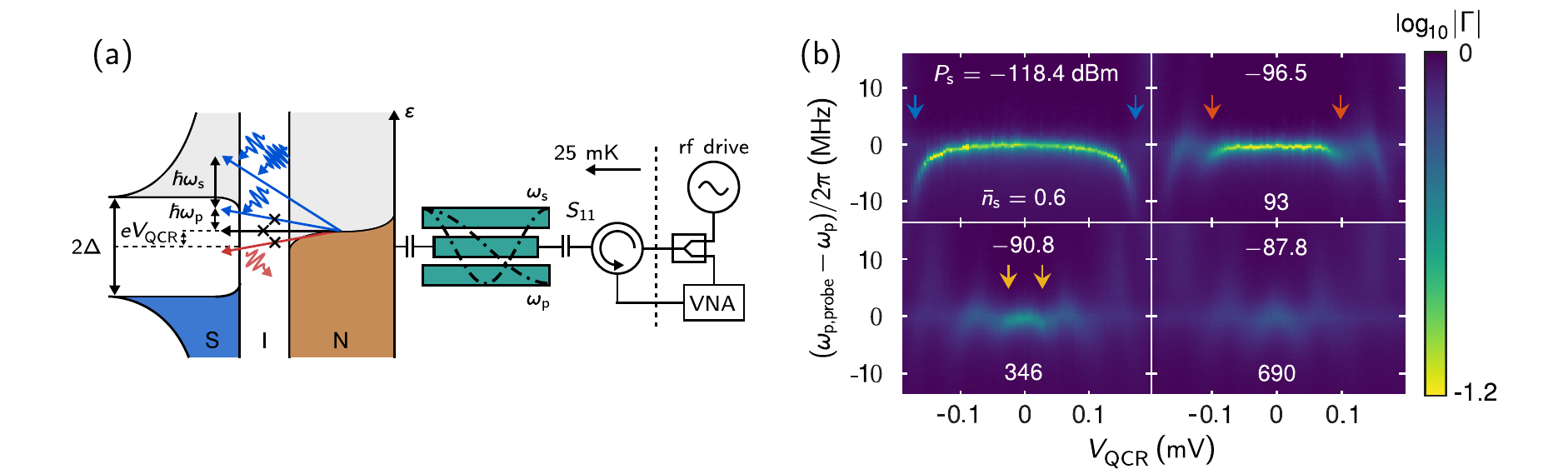}
  \caption{Radio-frequency quantum-circuit refrigerator. a) Device and measurement scheme of a CPW resonator is capacitively coupled to the normal-metal lead of the QCR together with an energy diagram for the photon-assisted tunneling. The fundamental mode of the resonator, $\omega_\mathrm{p}\approx2\pi\times8.8$~GHz, is controlled by the QCR which is operated with a dc voltage bias or with the excited second mode, $\omega_\mathrm{s}\approx2\pi\times17.6$~GHz, which activates multiphoton-assisted tunneling processes. b) Reflection coefficient at the indicated supporting-mode drive powers and corresponding mean photon numbers as a function of the QCR bias voltage and probe frequency. The measurement shows that an active rf QCR enhances the dissipation rate of the fundamental mode and induces an oscillating photon-number-dependent effective Lamb shift. Adapted from \textit{Photon-number-dependent effective Lamb shift} (DOI: 10.1103/PhysRevResearch.3.033126) with permission \cite{Viitanen21}. Copyright 2021, American Physical Society.}
  \label{fig:rfqcr}
\end{figure}

Above, the QCR has been operated solely with a bias voltage applied over the NIS junctions. This method shifts the electrochemical potentials of the normal-metal and superconducting leads relative to each other by $eV_\mathrm{QCR}$, as shown in \textbf{Figure~\ref{fig:rfqcr}}a, until photon-assisted electron tunneling processes are activated by absorbing photons from a connected quantum circuit. An alternative operation principle is provided in Viitanen et al. (2021)~\cite{Viitanen21} by a radio-frequency drive to a supporting mode in the system, such as a higher mode of a CPW resonator. The QCR is then activated by multiphoton-tunneling processes which absorb photons from both the supporting rf mode and the primary mode which we intend to control with the QCR. In a typical case, the primary mode with angular frequency $\omega_\mathrm{p}$ is in a single-photon regime and provides only up to a single photon for the electron tunneling processes, while the rf drive with angular frequency $\omega_\mathrm{s}$ provides the rest of the photons $\ell_\mathrm{s}$ required for the electron to tunnel over the gap $\Delta$. The on-state of the rf QCR is in the regime $eV_\mathrm{QCR}+\hbar\omega_\mathrm{p}+\ell_\mathrm{s}\hbar\omega_\mathrm{s} \lesssim \Delta$. In a typical device, such as in the experiment shown in Figure~\ref{fig:rfqcr}a with $\Delta/h\sim50$~GHz and $\omega_\mathrm{p}\sim2\pi\times10$~GHz, a purely rf operated QCR requires multiphoton processes of several photons or high frequencies. Although the higher-order transitions are heavily suppressed, they can be activated by providing a high supporting-mode occupation with a strong drive amplitude. Alternatively, the resonators can be fabricated to have large impedances, which strengthens the interaction of the modes with the tunnel junctions.

As with a dc-operated QCR, the rf QCR provides several orders of magnitude tunability for the dissipation experienced by the coupled quantum circuit. The coupling strength is obtained similarly as for a single-mode QCR but by adding the second mode to the core Hamiltonian of the coupled circuit and using the photon-assisted tunneling theory developed in Ref.~\cite{Silveri2017}. 
We obtain~\cite{Viitanen21}
\begin{equation}\label{eq:gamma_RF}
    \gamma_{\mathrm{T}, \mathrm{p}}\left(V_\mathrm{QCR}, \bar{n}_{\mathrm{s}}\right) = 2\underbrace{\pi \alpha_{\mathrm{p}}^{2} \frac{Z_{\mathrm{p}}}{R_{\mathrm{T}}}}_\text{primary mode} \underbrace{\sum_{k, l} P_{k}\left(\bar{n}_{\mathrm{s}}\right)\left|M_{k l}^{(\mathrm{s})}\right|^{2}}_\text{supporting rf mode} \underbrace{\sum_{\ell_{\mathrm{p}}, \tau=\pm 1} \ell_{\mathrm{p}} F\left(\tau eV_\mathrm{QCR}+\ell_{\mathrm{p}} \hbar \omega_{\mathrm{p}}+\ell_{\mathrm{s}} \hbar \omega_{\mathrm{s}}-E_{\mathrm{N}}\right)}_\text{tunnel junctions},
\end{equation}
where $\alpha_\mathrm{p}$ takes into account the capacitances of the resonator modes, $Z_\mathrm{p}$ is the characteristic impedance of the primary mode, $R_\mathrm{T}$ is the tunneling resistance, $P_k$ is the occupation probability of the $k$th Fock state of the supporting mode given a mean supporting-mode occupation $\bar{n}_\mathrm{s}$, $\left|M_{kl}^{(s)}\right|^2$ is the transition matrix element of the supporting mode from eigenstate $k$ to $l$, $\ell_\mathrm{p(s)}$ is the number of photons absorbed from the primary (supporting) mode, $E_\mathrm{N}$ is the charging energy, $F$ is the normalized forward tunneling rate, and $\tau$ takes into account both tunneling directions. For detailed mathematical definitions, see Ref.~\cite{Viitanen21}. 

The rf QCR provides the linear resonator mode a hybrid bosonic-fermionic environment consisting of the broadband vacuum environment provided by the QCR and the supporting higher mode of the resonator. Coupling to this hybrid environment induces a photon-number-dependent effective Lamb shift of several MHz to the harmonic mode, as shown in Figure~\ref{fig:rfqcr}b and modeled with Equation~\eqref{eq:lambshift}, but with the rf QCR coupling strength. Peculiarly, the staggered onset of different multiphoton absorption processes induce a Lamb shift which oscillates as a function of the bias voltage. Importantly, operating the QCR purely with an rf drive causes negligible Lamb shift while maintaining a typical range of tunable dissipation. Furthermore, a purely rf controlled QCR allows omitting the dc bias drive line simplifying the device and protecting the quantum circuit from additional noise. Note that the rf excitation can be applied from a shared transmission line in a multiplexed fashion, depending on the device, as well as from the NIS junction lead typically reserved for voltage bias.

An alternative straightforward approach to the quantum model of an rf QCR is to include an ac component to a dc bias control, which yields results closely resembling those presented above~\cite{Hsu2021}. Furthermore, the ac control was shown to provide similar reset fidelity and time as the dc control.

\section{Quantum-Circuit Refrigerator as an Incoherent Photon Source}
\label{sec8}

While in most cooling applications QCRs are typically operated in the voltage range $eV_{\mathrm{QCR}}<2\Delta$ to achieve photon absorption by electron tunneling, the opposite effect can be used for photon creation. This requires a larger bias voltage $eV_{\mathrm{QCR}}>2\Delta$, at which the photon creation rate approaches the absorption rate, thus increasing the effective temperature of the QCR environment. In this range, elastic tunneling is the primary mechanism of current through the NIS junctions. As opposed to inelastic tunneling described in Sec.~\ref{sec3}, the dissipation of the resonator modes is not directly affected by elastic contributions.  \\

Generally, the output power of the resonator
\begin{equation}
    P_{\mathrm{RT}}=\frac{2C^2\hbar\omega_0^3Z_0}{L_{\mathrm{res}}c_{\mathrm{res}}} (\bar{n}_{\mathrm{res}}-\bar{n}_{\mathrm{TL}})
    \label{eq:power}
\end{equation}
is determined by the difference of average photon numbers in the transmission line $\bar{n}_{\mathrm{TL}}$ and in the resonator 
$\bar{n}_{\mathrm{res}}$. Furthermore, it depends on the coupling capacitance between resonator and transmission line $C$, the fundamental resonator frequency $\omega_0$, the transmission line impedance $Z_0$, as well as the resonator length $L_{\mathrm{res}}$ and the capacitance per unit length $c_{\mathrm{res}}$. Since the average photon number follows the Bose distribution $\bar{n}_{\mathrm{res}}=\big( \exp[  \hbar \omega_0/(k_\mathrm{B}T_{\mathrm{res}})]-1\big)^{-1}$ for a thermal resonator state, the equation above enables us to determine both the resonator temperature $T_{\mathrm{res}}$ and the average photon number from measurements of the output power. This has been shown in an experimental study by Masuda et al. \cite{Masuda2018}, confirming a direct transformation of electric energy into microwave photons with the total output power surpassing that of 2.5-K thermal radiation while maintaining subkelvin chip temperatures. 

\begin{figure}[h]
\centering
  \includegraphics[width=\linewidth]{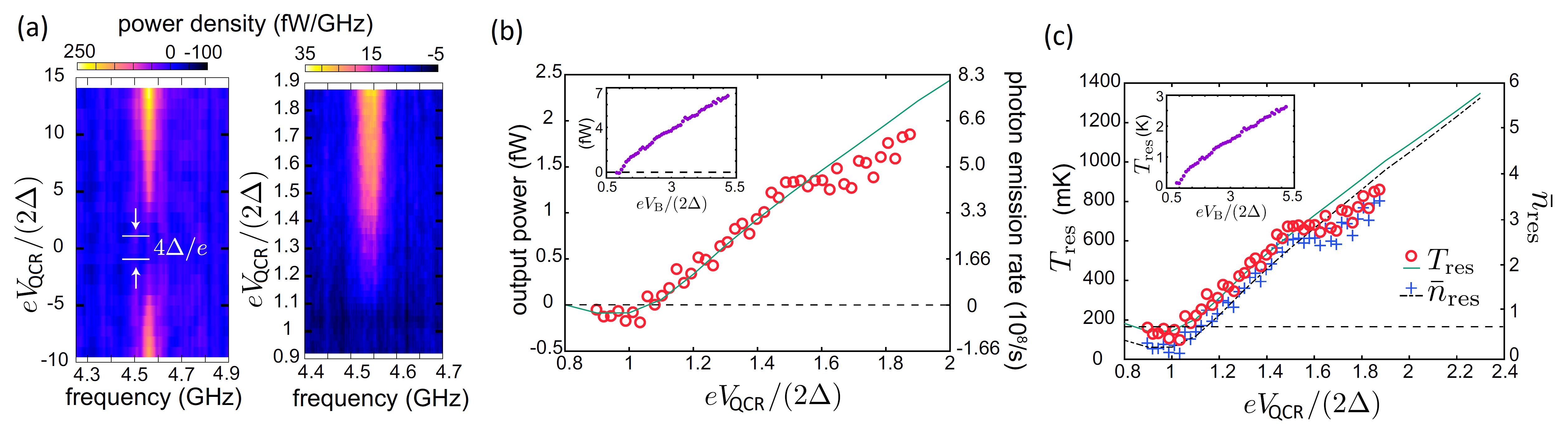}
  \caption{Output power and temperature measurements on a CPW resonator coupled to a QCR and NIS thermometer. a) Output power density as a function of frequency in a wide and narrow QCR bias voltage range. The highest power density is reached at maximum bias voltage in the vicinity of the resonator fundamental frequency $f_0 = 4.55\,\mathrm{GHz}$. The output vanishes as the bias voltage approaches the threshold $eV_{\mathrm{QCR}} = 2\Delta$. b) Frequency-integrated output power ($4.4-4.8\,\mathrm{GHz}$) from the measurement in a.  c) Temperature and average photon number in the resonator as calculated with equation \ref{eq:power} using the output power from  b. Adapted with permission \cite{Masuda2018}. Copyright 2018, Springer Nature.}
  \label{fig:Masuda}
\end{figure}

The measured output power spectrum in \textbf{Figure~\ref{fig:Masuda}}a shows maximum emission around the resonance frequency and negligible emission power at all frequencies for $|eV_{\mathrm{QCR}}|<2\Delta$. The integrated output power shows approximately linear behavior for high voltages, but a shallow dip around $eV_{\mathrm{QCR}} = 2\Delta$. This can be attributed to cooling of the resonator mode by the QCR. Note that the dip sustains in Figure~\ref{fig:Masuda}c where the measured output power has been converted into the resonator temperature and photon number. For $eV_{\mathrm{QCR}} = 10\Delta$, the resonator temperature reaches 2.5-K equivalent of thermal radiation.\\

This incoherent microwave source may have advantages over thermal resistor devices owing to its extensive and quick tunability, higher effective temperature, and reduction in excess heating in a similar fashion to shot-noise sources~\cite{Spietz2003, Gabelli2013}. Its design as an on-chip microwave source further eliminates unwanted losses from external circuitry. Due to its fixed frequency, it is an ideal candidate for tailored calibration purposes, e.g. for cryogenic radiation detectors~\cite{Clark2005, Spietz2003,kokkoniemi_bolometer_2020,kokkoniemi_nanobolometer_2019}.

\section{Calibration of Cryogenic Amplification Chains}
\label{sec9}

\begin{figure}[h]
  \includegraphics[width=\linewidth]{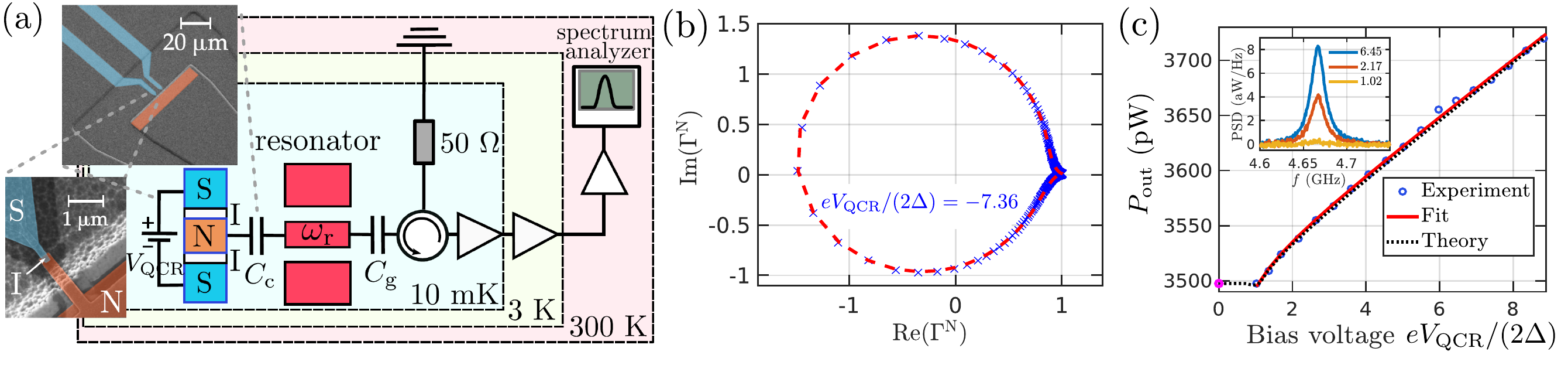}
  \caption{Method for characterizing the gain and noise temperature of an amplification chain. a) Schematic illustration of the sample design and the measurement setup. The sample incorporates a superconducting coplanar waveguide resonator that is capacitively coupled to a voltage-biased QCR and an output transmission line. The bias voltage of the QCR controls the rf power flowing from the source to the transmission line that is connected to the partly cryogenic amplification chain under study. Insets show false-color scanning-electron-microscope images of the sample in the vicinity of the NIS junctions. b) Example of the normalized voltage reflection coefficient $\Gamma^\textrm{N}$ (blue crosses) and its fit (red dashed line) that is used to estimate the damping rates $\gamma_\mathrm{tr}$, $\gamma_\mathrm{T}$, and $\gamma_\mathrm{x}$ for the given bias voltage. c) The measured output power as a function of the dc bias voltage of the QCR (blue circles), a fit of the equation $P_\mathrm{out} = aV_\mathrm{QCR} + b + c/V_\mathrm{QCR}$ (red line), and a prediction by the full theoretical model (dashed black line). The inset shows the increase in the measured power spectral density (PSD) for three bias voltages with respect to the power level at $eV_\mathrm{QCR}/(2\Delta) = 0$. Figures reproduced with permission \cite{hyyppa2019calibration}. Copyright 2019, AIP publishing.}
  \label{fig: gain calib}
\end{figure}

As discussed in the previous section, a QCR coupled to a superconducting resonator acts as a well-defined and controllable microwave photon source when the QCR is operated in the high-bias-voltage regime. The voltage-enabled control of the microwave power renders it feasible to use such a system for characterizing the total gain and noise temperature of an amplification chain connected to the QCR-resonator system. This was experimentally studied in Hyyppä et al. in 2019~\cite{hyyppa2019calibration} using a sample illustrated schematically in \textbf{Figure~\ref{fig: gain calib}}a. The sample incorporates a superconducting CPW resonator that is capacitively coupled to a dc-biased QCR and a transmission line. The transmission line is further connected to the amplification chain that is to be characterized at the frequency of the resonator. The characterization of the amplification chain can be carried out in two steps: First, one conducts standard voltage reflection measurements at different bias voltages and estimates the damping rates associated with dissipative reservoirs connected to the resonator as illustrated in  Figure~\ref{fig: gain calib}b. As the second step, one measures the power at the output of the amplification chain with a spectrum analyzer for varying bias voltages of the QCR. By fitting an analytic equation to the output power, it is possible to estimate the gain and noise temperature of the amplification chain as we discuss below.

The characterization protocol relies on the known relation between the input power of the transmission line $P_\mathrm{tr}$ and the bias voltage of the QCR.  In the high voltage regime $e V_\mathrm{QCR}/(2\Delta) \gg 1$, an accurate approximation for the the input power $P_\mathrm{tr}$ is obtained as \cite{hyyppa2019calibration}
\begin{equation}
P_\mathrm{tr} \approx \frac{\gtr \gbar}{\gtr + \gbar + \gint} \bigg \{ \frac{eV_\mathrm{QCR}}{4 } + \hbar \omr \bigg [ \frac{\gint (\Nint - \Ntr)}{\gbar} 
-N_\mathrm{tr} - \frac{1}{2} \bigg] - \frac{1}{2} \frac{\Delta^2}{eV_\mathrm{QCR}} \bigg( 1 + \frac{\gbar}{\gbar + \gtr + \gint} \bigg) \bigg \}, \label{eq: Ptr}
\end{equation}
where $\omr$ is the resonator frequency, $\gtr$ and $\Ntr$ denote the damping rate and effective photon number of the transmission line, $\gbar$ denotes the asymptotic damping rate of the QCR as $eV_\mathrm{QCR}/(2\Delta) \rightarrow \infty$, and $\gint$ and $\Nint$ denote the damping rate and effective photon number of excess losses. The power at the output of the amplification chain is given by $P_\mathrm{out} = G P_\mathrm{tr} + P_\mathrm{noise}$, where $G$ is the total gain of the amplification chain at the resonator frequency and $P_\mathrm{noise}$ is the power related to the effective noise temperature of the amplification chain. Therefore, the total gain $G$ can be estimated by fitting the equation  $P_\mathrm{out}(V_\mathrm{QCR}) = aV_\mathrm{QCR} + b + c/V_\mathrm{QCR}$ to the measured output power as a function of the bias voltage  as shown in Figure~\ref{fig: gain calib}c. As a result, the total gain of the amplification chain can be obtained as
\begin{equation}
G = \frac{4 a}{e} \frac{\gbar +\gtr + \gint }{\gbar \gtr}, \label{eq: gain}
\end{equation}
where $\gbar$, $\gtr$, and $\gint$ are known from the reflection measurements. Since the input power to the transmission line is vanishingly small for $eV_\mathrm{QCR}/(2\Delta) = 0$, one can further estimate the effective noise temperature of the amplification chain as 
\begin{equation}
T_\mathrm{noise} = \frac{P_\mathrm{out}(0)}{G k_\mathrm{B} \Delta f},  \label{eq: noise temp}
\end{equation}
where $\Delta f$ is the integration bandwidth used to measure the power with the spectrum analyzer. We note that the frequency range for the gain calibration may be expanded by rendering the frequency of the resonator tunable. This can be achieved, e.g., by placing a superconducting quantum interference device (SQUID)~\cite{partanen_exceptional_2019,sandberg2008tuning} or an array of SQUIDs into the resonator as done in Sec.~\ref{sec6} to be able to tune a two-resonator system into an exceptional point.

\section{Theory of Quantum-Circuit Refrigeration} \label{sec10}

In the previous sections we have reviewed a number of experiments with the QCR and discussed their successful comparison to theoretical predictions. Here, we provide a brief overview of the theoretical modelling of the QCR which has been developed in detail in Refs.~\cite{Silveri2017,Hsu2020,Hsu2021}. The starting point for modelling the system schematized in Figure~\ref{fig:QCR_circuit}a is the Hamiltonian description of the lumped-element circuit, the microscopic tunneling, and the electronic leads, given by
\begin{equation}
    \hat H_{\rm tot}=\underbrace{\hat H_{\rm C}+\hat H_\phi}_{\rm circuit}+\underbrace{\hat H_{\rm T}+\hat H_{\rm N}+\hat H_{\rm S}}_{\rm tunneling+leads},
\end{equation}
where the circuit Hamiltonian contains the capacitive $\hat H_{\rm C}$ and inductive $\hat H_{\phi}$ energy parts of the QCR and the quantum circuit. Importantly, the charge of the QCR and of the quantum circuit are coupled through the coupling capacitor $C_{\rm c}$. In other words, a change in the charge state of the QCR also induces a charge shift at the quantum circuit. The charge state of the QCR changes owing to quasiparticle tunneling between the normal-metal and the superconductor leads described by the tunneling Hamiltonian
\begin{equation}
    \hat H_{\rm T}=\sum_{k \ell \sigma}\left \{ T_{lk}\hat d^\dag_{l\sigma}\hat c_{k\sigma}\exp\left[-\textrm{i} \frac{e}{\hbar}\left(\alpha\hat \Phi_0+\hat\Phi_{\rm N}\right)\right] +T_{lk}\hat d_{l\sigma}\hat c^\dag_{k\sigma}\exp\left[\textrm{i} \frac{e}{\hbar}\left(\alpha\hat\Phi_0+\hat\Phi_{\rm N}\right)\right] \right\},
\end{equation}
where tunneling with the matrix elements $\sum_{lk\sigma}T_{lk}\propto 1/R_{\rm T}$ from the superconducting lead with annihilation operators $\hat c_{k\sigma}$ to the normal metal with creation operators $\hat d^\dag_{\ell \sigma}$ changes the charge of the normal-metal island $\hat Q_{\rm N}$ by $e$ and of the quantum circuit $\hat Q_0$ by $\alpha e$ [$\alpha=C_{\rm c}/(C_{\rm c}+C_{\rm j}+C_{\rm m})$] according to the displacement operator $\textrm{e}^{-\textrm{i}e\left(\alpha\hat \Phi_0+\hat\Phi_{\rm N}\right)/\hbar}$, which can cause state transitions in the quantum circuit. The energy of the microscopic electronic state of the leads and the bias shift by $eV$ are included in the Hamiltonians $\hat H_{\rm N}$ and $\hat H_{\rm S}$.

The calculation of the physical quantities of interest, such as the transition rates between the $m$:th and $m'$:th eigenstates of the quantum circuit $\Gamma_{m,m'}(V)$, the effective coupling strength $\gamma(V)=\Gamma_{1,0}(V)-\Gamma_{0,1}(V)$ or the effective QCR temperature $T_\textrm{T}(V)$, is based on treating the tunneling part $\hat H_\textrm{T}$ as a weak perturbation. The tunneling processes that exchange an energy quantum with the quantum circuit lead to relaxation or excitation of the circuit. 

The simplest approach for computing the transition rates for these photon-assisted processes is to use Fermi's golden rule and average over the electronic degrees of freedom assuming that them, as well as the excess charge on the normal island, can be effectively considered to be in thermal equilibrium~\cite{Silveri2017, Hsu2020}. In particular we assign a temperature $T_\textrm{N}$ to the electronic subsystems 
and a temperature $T_\textrm{Q}$ to the excess charge. The temperature $T_\textrm{N}$ equals for our purposes the base temperature of the fridge. At zero bias, we have $T_\textrm{Q} = T_\textrm{N}$, since in this case, there is no source of non-equilibrium, but in general $T_\textrm{Q}$ depends on the voltage in a non-monotonic way~\cite{Silveri2017}: in addition to the zero-bias minimum, another minimum is present at bias of order of, but smaller than $\Delta/e$ where $T_\textrm{Q}\simeq T_\textrm{N}/2$~\cite{Hsu2021}. This dependence raises the question if the assumption of effectively equilibrium charge distribution is correct when operating the QCR; this turns out to be the case, since over the short time in which the QCR is activated to refrigerate the system, the charge distribution is not strongly affected, as shown using a master equation approach in Ref.~\cite{Hsu2021}.

In using the QCR, we consider the following two important figures of merit: (i) the effective coupling strength $\gamma(V)$ between the QCR and the quantum circuit [cf. Equation~(\ref{eq:gamma_RF})] and (ii) the effective temperature  $T_\textrm{T}(V)$ to which the quantum circuit can be refrigerated. The coupling strength gives the time scale $\gamma^{-1}$ in which the effective temperature is reached in the quantum circuit. As shown in Figures~\ref{fig:lambshift}c,d and~\ref{fig:pulse_qcr_result}b, the coupling strength $\gamma$ varies between a minimum value $\gamma(V_{\rm off})$ at $$V_{\rm off}=0,$$ and a much larger value $\gamma(V_\mathrm{on})$ at a bias $V_\mathrm{on}$ somewhat below $\Delta/e$. The on-off ratio 
$$\gamma(V_\mathrm{on})/\gamma(V_\mathrm{off}) \approx \sqrt{\Delta/(\hbar\omega_{01})}/\gamma_{\rm D},$$ 
scales inversely with the so-called Dynes parameter $\gamma_{\rm D}$ and depends only weakly on the ratio between the superconductor gap and the lowest transition angular frequency of the circuit $\omega_{01}$. That is, the on-off ratio is set by the density of states within the superconducting gap, as characterized by $\gamma_{\rm D}$ which vanishes in the case of an ideal superconductor but achieves in typical experiments values down to $10^{-6}-10^{-4}$. This is natural since the quantum-circuit refrigeration is based on energy filtering in the tunneling processes set by the highly non-linear superconducting density of states, see Figure~\ref{fig:QCR_circuit}a.

The effective refrigeration temperature $T_\textrm{T}(V)$ coincides with the electron temperature $T_\textrm{N}$ at zero bias, changes non-monotonically with $V$, and reaches the minimum at an optimal bias 
$$|V_{\rm on}| \approx \Delta-\hbar \omega_{01}.$$ 
The optimal operational bias $V_\mathrm{on}$ and the achievable value of the effective temperature $T_\textrm{T}$ depend on the normal-metal electron temperature $T_\textrm{N}$. If the electron temperature $T_\textrm{N}$ is large compared to $\hbar\omega_{01}/k_\textrm{B}$, the effective temperature is half of that of the electron temperature $T_\textrm{T}(V_\mathrm{on})\approx T_\textrm{N}/2$. However, in the opposite regime of very low electron temperature $T_\textrm{N}$, the effective temperature of the QCR $T_\textrm{T}(V_\mathrm{on})$ may be only a fraction of $\hbar\omega_{01}/k_\textrm{B}$ due to the presence and typical dominance of the subgap tunneling determined by the finite Dynes parameter~\cite{Hsu2020}.


For a QCR coupled to a superconducting qubit such as a transmon or a capacitively-shunted flux-qubit, it has been theoretically obtained that with experimentally relevant qubit and QCR parameters, one can reach reset infidelities $1-\mathcal{F}_{\rm r}\approx 10^{-5}-10^{-3}$ in  $10$--$100$~ns depending on the achievable electron temperature $T_{\rm N}$~\cite{Hsu2020}. If a QCR is directly coupled to a superconducting qubit, the tunneling events and the associated charge fluctuations may cause additional dephasing to the qubit owing to the charge dispersion of the energy levels. Due to the subgap density of states, tunneling is also present in the off-state of the QCR. In practice, however, this effect can be minimized below the level of the natural qubit dephasing rate by taking the ratio of the Josephson energy and the charging energy to be sufficiently large, $E_{\rm J}/E_{\rm C}\lesssim 100$. 

In the latest theoretical work, we have studied a QCR where the size of the normal-metal island is reduced to a limit where the charging energy of the normal-metal island and the spacing of the electronic energy levels dominate over other energy scales~\cite{Hsu2021}. In this case, with a suitably chosen bias, the quantum dot QCR can realize negative damping, that is, microwave gain, or produce a non-classical state in a coupled resonator. In practice, this limit may be realized by voltage-biased superconductor--insulator--quantum-dot--insulator--superconductor junctions (SIQDIS)~\cite{bruhat_cavity_2016, van_zanten_single_2016,van_zanten_probing_2015}.


\section{Conclusion and Outlook}\label{sec11}
As outlined above, the QCR has proven a valuable component in circuit quantum electrodynamics. Its discovery was reported in experiments aiming to extend previous observations of quantum-limited heat conduction by microwave photons from transmission lines to resonators, where accidentally, instead of ohmic losses, the photon-assisted tunneling was dominating. 
Coupled to a resonator, the QCR has established an experimental method for efficient cooling and control of dissipation. Further studies have discovered a Lamb shift of the resonator frequency, both for dc and rf operation. In the high-voltage operation regime, the QCR has been successfully utilized to generate incoherent microwave photons, and to characterize the gain and noise temperature of cryogenic amplification chains.

Its versatility as a stand-alone, on-chip device, combined with its straightforward integration in superconducting circuits renders the QCR a prime candidate for providing tunable dissipation in superconducting qubits~\cite{Hsu2020,Yoshioka2021}. In this world of increasing qubit lifetimes, for transmon qubits currently at the level of $500\,\mathrm{\mu s}$~\cite{wang_transmon2021}, the need of on-demand active qubit reset will only increase in the future~\cite{divincenzo_physical_2000}. 

Theoretical calculations by Hsu et al.~\cite{Hsu2020} have shown that a QCR may decrease the excited-state population of a qubit by an order of magnitude in $1.5\,\mathrm{ns}$ with a minimum reset infidelity of $5\times10^{-5}$ at an electron temperature of 10 mK. Although the QCR is a dissipative element, it induces no significant effect on gate or readout errors in its off-state~\cite{Hsu2020}. Yet, the qubit can be further protected by coupling the QCR to the qubit via a resonator~\cite{Yoshioka2021}. The qubit reset may be carried out by moving the qubit occupation to the resonator which is reset by the QCR. 

In the reset of superconducting qubits, the QCR has a key advantage: the natural depletion of highly excited qubit states and a possible resonator. This is not necessarily the case for protocols based on parametric modulation, which only transfer the qubit excitation into a coupled resonator, thus increasing the reset time by resonator decay~\cite{Zhou2021}. 

The next major experimental goal is to show a high-fidelity qubit reset using the QCR. This also allows one to accurately extract the effective temperature of the QCR environment and hence to optimize it. Thus it is likely that the QCR design will evolve. Possibly lower Dynes parameters can be achieved by improved filtering out of high-frequency radiation arriving at the QCR and the superconductors in its vicinity. Intriguingly, the QCR can be utilized in applications beyond reset such as dissipation-driven quantum state engineering \cite{verstraete_quantum_2009} and for ground-state population control in quantum annealing \cite{Johnson2011, Dickson2013}. Thus it seems that we have thus far witnessed only the beginning of the story of quantum-circuit refrigeration.


\medskip

\medskip
\textbf{Acknowledgements} \par 
This work was financially supported by the Finnish Cultural Foundation, the Academy of Finland under Grant No.~318937 and under its Centres of Excellence Program (project No.~312300), and the European Research Council under Grant No. 681311 (QUESS).

\medskip
\textbf{Conflict of Interest} \par 
The authors declare that M.M. is a Co-Founder of IQM.

\medskip

%
\bibliographystyle{MSP}
\bibliography{paper}

\newpage

%





\end{document}